# Adaptive ensemble simulations of biomolecules

Peter M. Kasson[a,b] and Shantenu Jha[c,d]

[a] Departments of Molecular Physiology and of Biomedical Engineering, University of Virginia, Charlottesville VA 22908 United States.

[b] Science for Life Laboratory, Department of Cell and Molecular Biology, Uppsala University, Uppsala 75146 Sweden.

[c] Department of Electrical and Computer Engineering, Rutgers University, Piscataway NJ 08854 United States.

[d] Center for Data-Driven Discovery, Brookhaven National Laboratory, Upton NY 11793 United States.

Correspondence may be directed to kasson@virginia.edu or shantenu.jha@rutgers.edu

Recent advances in both theory and computational power have created opportunities to simulate biomolecular processes more efficiently using adaptive ensemble simulations. Ensemble simulations are now widely used to compute a number of individual simulation trajectories and analyze statistics across them. Adaptive ensemble simulations offer a further level of sophistication and flexibility by enabling high-level algorithms to control simulations based on intermediate results. We review some of the adaptive ensemble algorithms and software infrastructure currently in use and outline where the complexities of implementing adaptive simulation have limited algorithmic innovation to date. We describe an adaptive ensemble API to overcome some of these barriers and more flexibly and simply express adaptive simulation algorithms to help realize the power of this type of simulation.

## Introduction

In recent years, molecular dynamics simulation has been increasingly utilized for quantitative prediction of and insight into biophysical problems. Moving beyond visualization and qualitative insight, robust statistical estimation of thermodynamic, kinetic, or structural properties of biomolecules is now within reach. However, this new horizon creates a set of challenges, since statistical estimation of these quantities requires many simulated "observations" of the desired process and quantitative insight thus comes at a cost of substantial computational requirements. Much like single-molecule experiments, the most efficient way to obtain these simulated observations is using collections or *ensembles* of simulations rather than a single extremely long trajectory of single-event observations [1,2]. Such ensemble approaches have also been used to predict effects of mutations at different sites on a protein, to estimate transition states and free-energy barriers, and make other similar quantitative predictions [3-10].

A further advance in the range of biomolecular processes that can be feasibly simulated comes from not only simulating biomolecular ensembles but running these simulations in an *adaptive* manner, where high-level algorithms are used to determine the next round of simulations based on the results of the previous one. Such adaptive algorithms can increase simulation efficiency by greater than a thousand-fold [11-13] but require a more sophisticated software infrastructure to support them. Here, we review some of the biophysical insights gained via ensemble simulations, the software needs and current capabilities for flexibly and efficiently running such calculations, and a pathway to filling some of the unmet needs in this area.

## Adaptive ensemble methods for simulating biomolecules

Although frameworks for adaptive ensemble simulation have been developed only recently [12,14,15], simpler algorithms for adaptive ensemble simulation have been in use for many years. Many algorithms pre-specify the sequence of computational simulations, but the results of each set of simulations are used to determine the inputs for the next round. Algorithms where not only the simulation parameters but even the type of computational operation to perform depends on intermediate results are rarer, due in large part to the higher barrier to implementing such algorithms, but these present perhaps the most exciting and powerful set of simulation approaches.

Replica exchange molecular dynamics is a long-standing and widely used ensemble method where individual simulations within an ensemble exchange coordinates over the course of the simulation. Depending on the exchange algorithm, this can be performed in either a non-adaptive or an adaptive fashion. Replica exchange was originally formulated as temperature replica exchange [16-18], where an ensemble of simulations is run at different temperatures to facilitate escape from energy minima, exchanging coordinates via a Monte Carlo criterion. In a related method, generalized-ensemble simulation, exchange can be performed over larger numbers of generalized "coordinates", including between different Hamiltonians, and different exchange algorithms between ensemble members can be employed [19-21]. This has permitted exploration of free-energy surfaces that are less accessible to temperature replica exchange alone. One example of increased adaptivity in such simulations comes from adaptive placement of scaling parameters ("lambda values") in free-energy perturbation and similar calculations to optimize statistical convergence [22-27]. Expanded-ensemble simulations are related to replica exchange; in terms of parallelization, they can be seen as serial adaptive algorithms that can be parallelized whereas replica exchange is an ensemble algorithm that can be made adaptive. Adaptivity in these cases has largely been supported via explicit implementation in molecular dynamics software packages, and a more flexible platform for such adaptive algorithms would potentiate further algorithmic development, reuse of existing algorithms by other scientists, and resulting scientific progress.  Conversely, metadynamics approaches have been implemented primarily in high-level software such as PLUMED [28,29] that abstracts the adaptivity for algorithms such as multiple-walker metadynamics [30] but requires explicit job scheduling.

Adaptive ensemble simulation has been particularly helpful in biomolecular simulation algorithms where each individual simulation uses an identical unbiased Hamiltonian but where placement of simulations in phase space is optimized to improve estimation of the kinetics and thermodynamics of a biomolecular process.  In an ensemble formulation, placement of unbiased trajectories in phase space involving choosing which trajectories to extend or from which already-sampled points in position and velocity space to start new trajectories.  Some of these unbiased-trajectory algorithms include milestoning, weighted-ensemble simulation, and related techniques [15,31-34]; each of these has been implemented in custom software packages to facilitate the adaptive logic and post-processing.

Methods to construct Markov State Models from molecular dynamics simulations provide a similar set of powerful approaches for analyzing molecular kinetics using unbiased individual trajectories [35-38].  The choice of starting points for these trajectories can be optimized to reduce the uncertainty of the resulting model: it has been demonstrated retrospectively and then prospectively that adaptive sampling with Markov State Models increases convergence efficiency by several orders of magnitude.  Adaptive sampling methods have recently been applied with great success to complex biomolecular processes [39]. Another recent study combines biased umbrella sampling simulations with Markov State Model-inspired estimators and adaptive sampling, showing how the facile combination of methods can potentiate further insight [40].  However, this can be difficult because most implementations of such methods have been in special-purpose code.  One exception is Copernicus [12,27], but that has other limitations as detailed below.

**Designing software systems for adaptive ensemble methods**
The broad range of adaptive ensemble simulation algorithms impose similarly diverse requirements on the underlying software infrastructure. Algorithms differ in the frequency of communication between ensemble members, local versus non-local communication, and the type of information exchanged. Two adaptive simulation work/data flow diagrams are schematized in Figure 1.  Adaptive changes can alter the number of tasks being performed (how many ensemble members in a simulation), the parameters of those tasks (placement of temperature or lambda values in an expanded-ensemble simulation), or even which tasks are being performed when (e.g. branching between simulations to converge a bound-complex ensemble and free-energy-perturbation simulations to measure binding of a new candidate ligand and either accept or reject that ligand for inclusion into the main simulation loop).  The logic to specify such changes can rely on a single simulation within an ensemble, an operation across an ensemble, or even external criteria, such as changes in resource availability or new experimental data.

Despite this diversity, a key commonality among adaptive algorithms is that they can be expressed at a high level, such that the adaptive logic itself is independent of simulation details. This separation of adaptive operations from simulation internals provides a useful and important abstraction for both methods developers and the software system. Adaptive operations that are expressed independent of the internal details of tasks facilitate MD software package agnosticism and simpler expression of different types of adaptivity and responses to adaptivity.

This promotes facile development of new methods while facilitating optimization and performance engineering that will be needed at large scales.

Expressing adaptive algorithms in this more abstract manner, as computational processes separate from but operating on independent ensemble members, creates several implementation challenges. These include coordination and consistency across distributed execution components, scalable communication between independent simulations and efficient stop and restart of simulations. Separating the adaptive logic from underlying execution management software allows the complexity to be contained within the internal implementation of the software system and not be exposed to the user. This approach also enables transparent low-level optimization and adjustment to fluctuations in workload and resource availability. We believe sophisticated runtime systems will be necessary to support adaptive ensemble algorithms at scale, as similar runtime management has been required for efficient execution of even relatively static ensemble workloads at scale [4,41,42] (Figure 2). It has been well known that on MapReduce and similar parallel architectures, completion of a few "lagging" tasks in an ensemble dominates the overall time to completion [43]. Although advanced runtime systems can mitigate this problem, asynchronous analysis tasks can algorithmically bypass it.

**Steps toward greater adaptivity – State of the art**
Several software systems have been used for adaptive ensemble methods [12,27,44,45]. Most solutions fall into one of two categories: monolithic general-purpose workflow systems that do not have "native" support for adaptive algorithms, or where adaptive algorithms are embedded internal to the MD engine/package [46-50]. Relatively few support ensembles of tasks or adaptive operations as first-class entities. Most workflow systems support adaptation as a response to fault tolerance [51] rather than adaptive logic based on intermediate results. Conversely, many biomolecular simulation packages (AMBER, Gromacs, CHARMM, and NAMD [46-50]) provide some specific ensemble or adaptive capabilities. However, these are tightly coupled to the code of the MD packages, and implemented in a manner such that it is not easy for users to add new adaptive algorithms.

A smaller number of "advanced workflow" packages or dataflow programming languages offer a greater degree of adaptivity and are usable for molecular simulations. Scalable ensemble-based adaptive algorithms require support at multiple levels: programming models and APIs, execution models and runtime system etc. In addition to programming and execution model choices, there are open questions about the granularity of tasks and suitable abstractions to express adaptivity. Swift/T [44,52] and Copernicus [12,27] are prominent examples of data-driven task parallelism that support adaptive applications, but they differ significantly in their programming model and how they support adaptivity. Swift/T is primarily designed to extract parallelism from scripts that express data dependencies between instances of existing applications. The Swift script is compiled to run the sequential or parallel applications within an MPI application using a sophisticated runtime system to support the execution of many tasks. It has to date not been used for adaptive biomolecular simulations. Copernicus' [12,27] data-driven execution model considers individual (MD) simulations as the unit of execution (i.e., task) and adaptivity managed by modifying the task-graph. In both of these packages, as in our proposed formalism, operations are executed when their inputs are satisfied. This greatly simplifies parallelization, as parallel execution does not have to be explicitly specified but results naturally from a lack of data dependencies.

Fireworks [53] is another ensemble workflow package used primarily in the materials simulation community that allows for dynamic changes to the workflow graph but has not been utilized for the adaptive simulations we describe here. Another software package, Ensemble Toolkit [54], has recently been extended to support some adaptive simulations (Markov State Models and expanded-ensemble simulations) [14]; these capabilities have also been applied to drug binding-affinity calculations [55]. However, we note that all of these packages have primarily been applied to adaptive calculations by the package developers or their collaborators, suggesting that flexibility and ease-of-use could be improved to facilitate broader uptake.

**An Adaptive Ensemble API**
In order to more flexibly and simply express adaptive simulation algorithms, we propose an Adaptive Ensemble API. This API could either be used directly in user code to specify and run adaptive simulations, be utilized by developers of new computational methods in their code, or be used for library calls within molecular dynamics software packages to more flexibly and powerfully implement ensemble simulations.

The following set of functions should be necessary and sufficient to express the required adaptivity for a broad range of ensemble methods in computational biophysics. All the functions operate on compute kernels--discrete computational tasks such as running an MD simulation. Core adaptive capability is provided by **while**() and **if**(). Each of these operations enables conditional execution of code paths depending on the results of some compute kernel (or API operation). The **map** and **reduce** operations provide basic parallel functionality similar to MapReduce but with the important difference that **reduce**() returns a variable-dimension output. Briefly, **map**() applies a compute kernel to an arrayed set of inputs in a parallel fashion. **Reduce**() takes an arrayed set of inputs and applies a compute kernel to the array, producing either a single output or an arrayed output. Together, these operations can be used to execute a wide variety of ensemble workloads in a parallel fashion. The **async**() and **cancel**() operations add capabilities for asynchronous tasks that can operate on intermediate outputs, for instance analysis processes that monitor simulation outputs and return decisions on whether to cancel them and spawn new simulations. **Async**() is critical to efficient ensemble computing because it enables non-blocking parallel operations and can be used to avoid waiting for a lagging task. Kernels are specified with inputs, outputs that are written once, and intermediate shared variables that can contain intermediate outputs or in-flight inputs. Figure 3 provides a listing of these operations and their syntax, while Figure 4 shows how an adaptive ensemble algorithm can be expressed using these API operations.

Such an API could be interfaced to existing ensemble packages either directly or via a task-graph manager to optimize execution, depending on the requirements and capabilities of the underlying software. In addition to a top-level interface for users and simulation methods developers, this API could be used by existing molecular dynamics packages (or APIs such as the ensemble APIs for GROMACS[56]) as a set of library calls to manage adaptive execution and facilitate the implementation of new adaptive simulation methods.

**Conclusions**
Adaptive ensemble simulation methods, from the simple to the complex, have already made a strong impact on biomolecular simulation and our understanding of biomolecular kinetics and thermodynamics. This is despite the relative lack of tools to easily express sophisticated adaptive algorithms and run them in a scalable fashion. As molecular simulations are used to address questions of increasing biological complexity, the gains in algorithmic sophistication and computational efficiency from adaptive ensemble methods will become critical in generating quantitative insight into biological problems. It is our hope that the availability of APIs such as the adaptive ensemble API we describe here will facilitate the expression of new, innovative adaptive algorithms and the implementation and comparison of these algorithms for many more simulation packages and many more biological problems of interest. Continued development of software infrastructure for adaptive ensemble simulations, new adaptive methods, and new applications of these methods to important biophysical and structural problems have the potential to greatly increase simulation's utility as a tool for quantitative, rather than only qualitative, biomolecular insight.

**Acknowledgements**
The authors thank Michael Shirts, Thomas Cheatham, Eric Irrgang, Anubhav Jain, and Daniel Katz for many helpful discussions. SJ also thanks Vivek Balasubramanian for prototypes and performance experiments that helped develop some of these ideas.

**Funding**
This work was supported by National Instutites of Health [grant R01 GM115790 to PMK] and National Science Foundation [grant 1547580 (Molecular Science Software Insitute) to SJ].

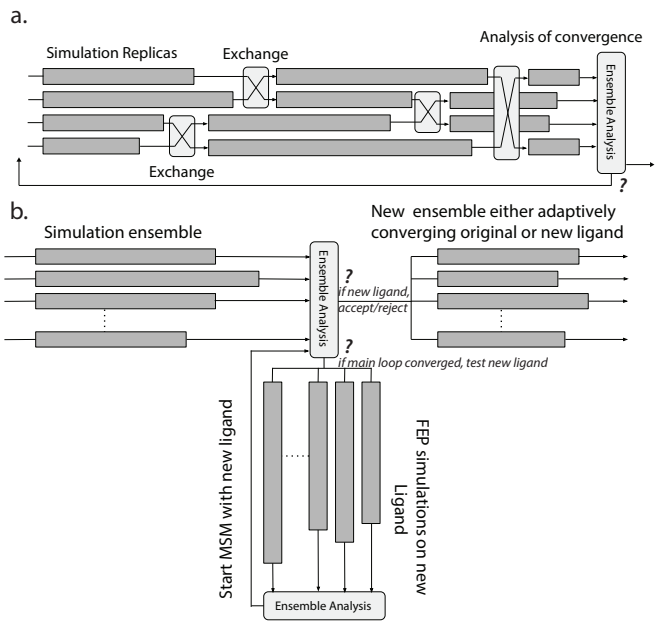

**Figure 1. Adaptive ensemble work diagrams**. Panel **a** schematizes an asynchronous replica exchange loop. Ensemble members are run asynchronously, so there is no global barrier before exchange or analysis. This is not *per se* an adaptive concern but is required for many efficient adaptive algorithms. An ensemble analysis then tests for convergence and either re-triggers the loop (perhaps with altered parameters) or writes a final output. Panel **b** schematizes more complex adaptive logic, where an initial simulation ensemble of protein-ligand interaction asynchronously triggers an analysis calculation (which could be clustering and Markov State Model construction). This analysis calculation either adaptively reseeds the ensemble simulation run or, if the run is converged, starts an ensemble free-energy-perturbation (FEP) calculation on a new ligand (lower branch). Depending on the result of this FEP calculation, it is either "accepted" and a new Markov State Model calculation started with the new ligand, or it is "rejected" and a new ligand tested. In all schemas, dark

gray rectangles indicate ensemble simulations and light gray rectangles indicate analyses.

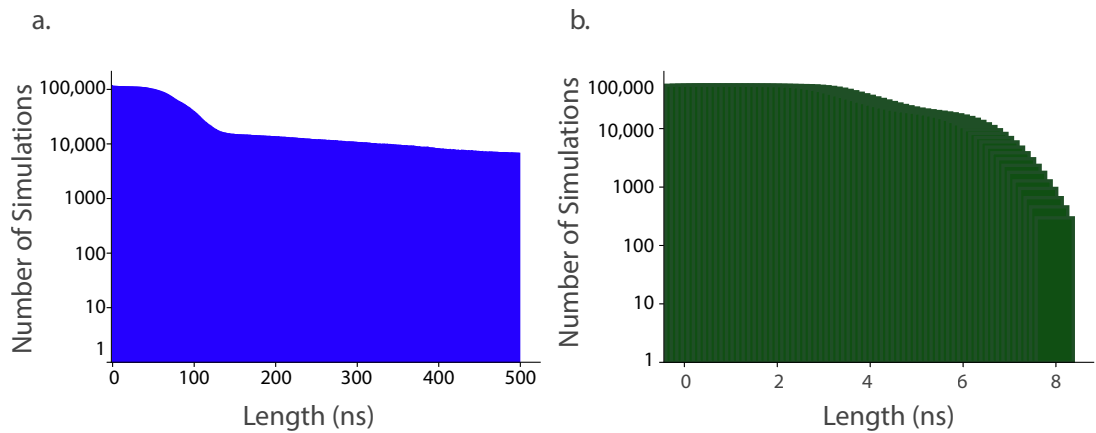

**Figure 2. Need for sophisticated runtime systems to manage adaptive ensemble simulations**. Survival curves of number of ensemble members versus simulation time reached are plotted for simulations run on Folding@Home (**a**) and Google Exacycle (**b**). Due to stochastic factors, large ensemble calculations show a near-exponential decay in number of ensemble members reaching a given simulation length. This is somewhat mitigated by ensemble management algorithms. This decay causes a "long tail" in simulation completion times, which can result in substantial inefficiencies if a global barrier exists such that all simulations must complete prior to analysis. This can be partially mitigated by advanced runtime systems, but asynchronous analyses that do not require all simulations to complete can algorithmically overcome this issue. Simulation data plotted from [4].

`map(f, inp)` - Run kernel f in parallel on each input member of array inp. Produces output of equal dimension to inp.
`reduce(f, inp)` - Run kernel f collectively on all of the members of array inp. This produces an output of variable dimension. This function is used for operations such as clustering.
`while(reduce(f, inp)): reduce(g, inp2)` - Reduce kernel f on array inp. While the result of this operation is true, reduce kernel g on array inp2.
`if(reduce(f, inp): run(g, inp2); else run(h, inp2)` - Reduce kernel f on array inp, if true reduce kernel g on inp2 and if not reduce kernel h on inp2.
`async [operation]` - specifies that the operation runs in the background and the operation can write intermediate as well as final outputs.
`cancel (operation handle, boolean mask)` - takes a mask of dimension equal to the parallelism of the operation handle and cancels the subtasks at locations where the mask value is True.

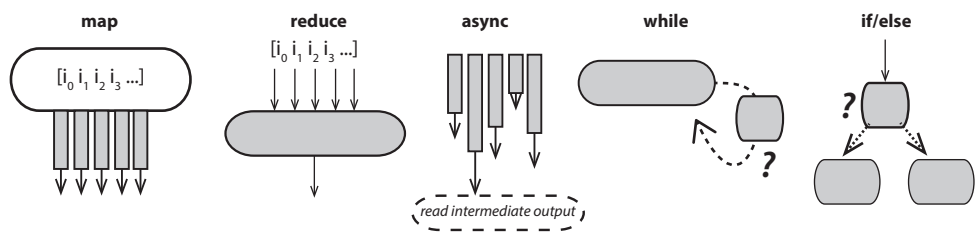

**Figure 3**. Operations comprising the Adaptive Ensemble API. We denote kernels as *f*, *g*, and *h*, and each kernel takes inputs, possibly arrayed, denoted **inp**. In the illustrative schemas, gray boxes indicate kernels being executed, solid lines with errors indicate inputs, dotted lines indicate logical flow, and question marks indicate branch points.

```
md = gmx.workflow.from_tpr_and_structures(tprfile,
    ensemble.input_placeholder()) # init simulations using placeholder
context = gmx.context.managedParallelContext(md)
# MD simulations to be managed by EnsembleAPI
# presumes an MSM class that can read Gromacs API output objects
msm = msmpackage.GromacsAdaptiveMSM()
a = map(md, siminputs)
b = reduce(msm.MSMsample, a.outputs)
while reduce(msm.isnotconverged(), b.outputs):
  c = async map(md, trajinputs)
  while not reduce(any_true, (b= reduce(msm.MSMsample,
      c.outputs)).outputs_stop):
    reduce(waitforincrement, c.outputs)
  cancel(c, b.outputs_stop)
  trajinputs = b.outputs
```

**Figure 4**. Pseudocode implementing adaptive Markov State Models using Adaptive Ensemble API. For demonstration purposes, we have shown an implementation using gmxapi [56]; the approach should easily generalize to other Python frontends for molecular simulation programs.